\documentclass{webofc}

\usepackage[varg]{txfonts}   
\usepackage{hyperref}
\usepackage{url}
\usepackage{float}
\usepackage[left]{lineno}
\usepackage{subcaption}
\hypersetup{colorlinks=true,citecolor=blue,urlcolor=blue,linkcolor=blue}
\usepackage{lineno, blindtext, indentfirst}
\begin{document}
%

\title{Production of Light Nuclei in Au+Au Collisions at \\$\sqrt{s_{\rm NN}} = 7.7-27$ GeV from STAR BES-II}
%
%

\author{\firstname{Yixuan} \lastname{Jin}\inst{1,2}\fnsep\thanks{\email{jyx@mails.ccnu.edu.cn}} (for the STAR Collaboration)}
\institute{Institute of Particle Physics and Key Laboratory of Quark \& Lepton Physics (MOE), \\Central China Normal University, 430079, Wuhan, China.
     \and Physics Institute, Heidelberg University, 69120, Heidelberg, Germany.}

\abstract{The studies of the production of light nuclei, such as deuteron and helium nuclei, in heavy-ion collisions are essential for understanding the dynamics of nuclear matter under extreme conditions. The yields and ratios of light nuclei serve as an effective method to distinguish between the thermal and coalescence models of light nuclei formation. Within the coalescence framework, the energy dependence of the coalescence parameters reflects the effective volume of the collision system, while in the thermal model yields are governed by chemical freeze-out conditions. The significantly larger datasets from the STAR Beam Energy Scan Phase II (BES-II), combined with enhanced detector capabilities, allow more precise and comprehensive measurements than phase I.

In these proceedings, we present measurements of light nuclei production, including p, $\rm \bar{p}$, d, $\rm \bar{d}$, $\rm ^3He$, in Au+Au collisions at BES-II energies of $\sqrt{s_{\rm NN}} = 7.7 - 27$ GeV. The results include centrality-dependent transverse momentum spectra and yields ($\mathrm{d}N/\mathrm{d}y$), along with coalescence parameters $B_\mathrm{A}$ and particle yield ratios. The physics implications of these results are discussed.
}
\maketitle

\section{Introduction}
\label{intro}
The Beam Energy Scan (BES) program at the Relativistic Heavy-Ion Collider (RHIC) is designed to investigate the phase structure of quantum chromodynamics (QCD) by varying the collision energy. The second phase of the program (BES-II), featuring enhanced detector capabilities and a tenfold increase in data compared to BES-I \cite{BESI_d, BESI_t, 3GeV}, allows for more precise measurements in the exploration of the quark-gluon plasma (QGP), the critical point, and possible first-order phase transition \cite{QCDPhase, Luo:2022mtp, Chen:2024aom}.

The production of light nuclei and their antiparticles in relativistic heavy-ion collisions has attracted sustained interest. While various models have been proposed to describe the production mechanism of light clusters, two main approaches are widely discussed. The thermal model assumes that the primordial particle yields are fixed at chemical freeze-out, when inelastic interactions cease. These yields are determined by thermodynamic parameters such as the chemical freeze-out temperature ($T_{\rm ch}$) and baryon chemical potential ($\mu_{\rm B}$), with final-state corrections from resonance decays \cite{ThermalCoalescenceTheory}. The coalescence approach, on the other hand, assumes that light nuclei are formed after kinetic freeze-out, when nucleons combine if they are close in both coordinate and momentum space. This process is sensitive to the source size \cite{Coalescence}.


\section{Signal Extraction and Corrections}
In light nuclei production analyses, two major detectors of the STAR experiment are utilized for charged particle reconstruction and identification. The Time Projection Chamber (TPC) provides precise measurements of the average ionization energy loss per unit length ($\langle \mathrm{d}E/\mathrm{d}x\rangle$) for each charged particle and, with the upgrade of the inner sectors, offers an extended momentum coverage at low momentum range. To identify particles at higher momenta, the barrel- and end-cap Time-of-Flight (TOF) detectors provide mass information.

The transverse momentum ($p_{\rm T}$) spectra are obtained after applying corrections for energy loss, TPC tracking efficiency, TOF matching efficiency, and knock-out effects—specifically for protons and deuterons. 
Feed-down contributions from weak decays of strange baryons to protons and anti-protons are subtracted via a data-driven approach, which uses the inclusive (anti-)proton and the yields of strange hadrons measured by the STAR experiment.
Point-to-point systematic uncertainties in $p_{\rm T}$ spectra are estimated by varying analysis cuts.

\section{Results}
\subsection{Corrected transverse momentum spectra}
\begin{figure}[H]
     \centering
     \includegraphics[width=0.55\textwidth]{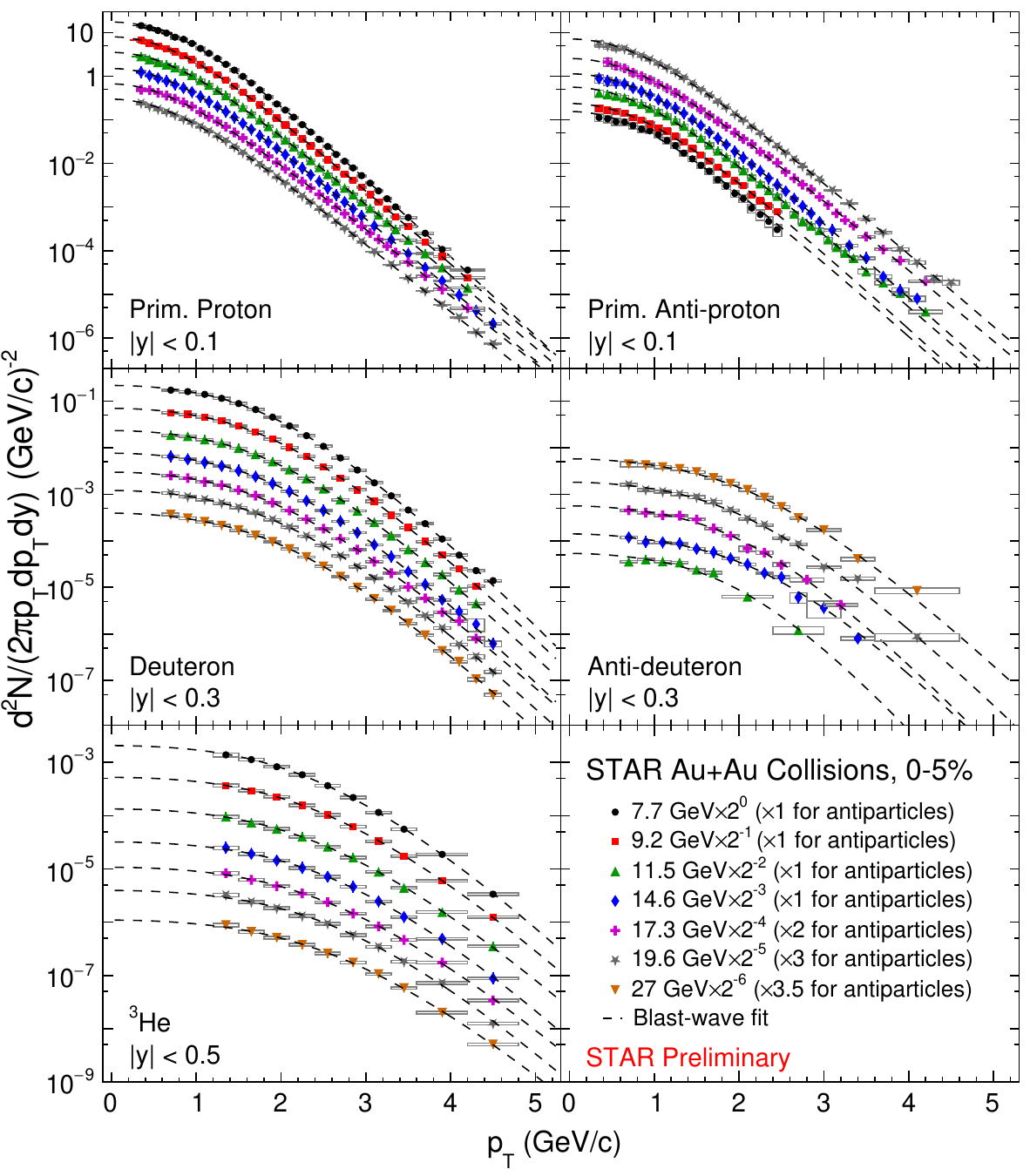}
     \caption{Transverse momentum spectra of light nuclei and protons in central ($0$--$5\%$) Au+Au collisions at mid-rapidity in the STAR BES-II program at $\sqrt{s_{\rm NN}} = 7.7 - 27$ GeV. The rapidity selections are $|y| < 0.1$ for (anti-)protons, $|y| < 0.3$ for (anti-)deuterons, and $|y| < 0.5$ for $\rm ^3He$. The dashed lines represent the corresponding Blast-Wave fits. Statistical and systematic uncertainties are shown as vertical lines and shaded boxes, respectively. The spectra are scaled for clarity.}
     \label{fig-2}
\end{figure}

The $p_{\rm T}$ spectra of light nuclei and protons in central Au+Au collisions from the STAR BES-II program are shown in Fig.~\ref{fig-2}. The extended $p_{\rm T}$ coverage, enabled by higher statistics and improved detectors, is achieved in BES-II, and the results are consistent with those from BES-I within uncertainties.

\subsection{Particle Yields ($\mathrm{d}N/\mathrm{d}y$) and Mean $p_{\rm T}$ ($\langle p_{\rm T} \rangle$)}
\begin{figure}[h]
     \centering
     \begin{subfigure}[b]{0.435\textwidth}
          \includegraphics[width=\textwidth]{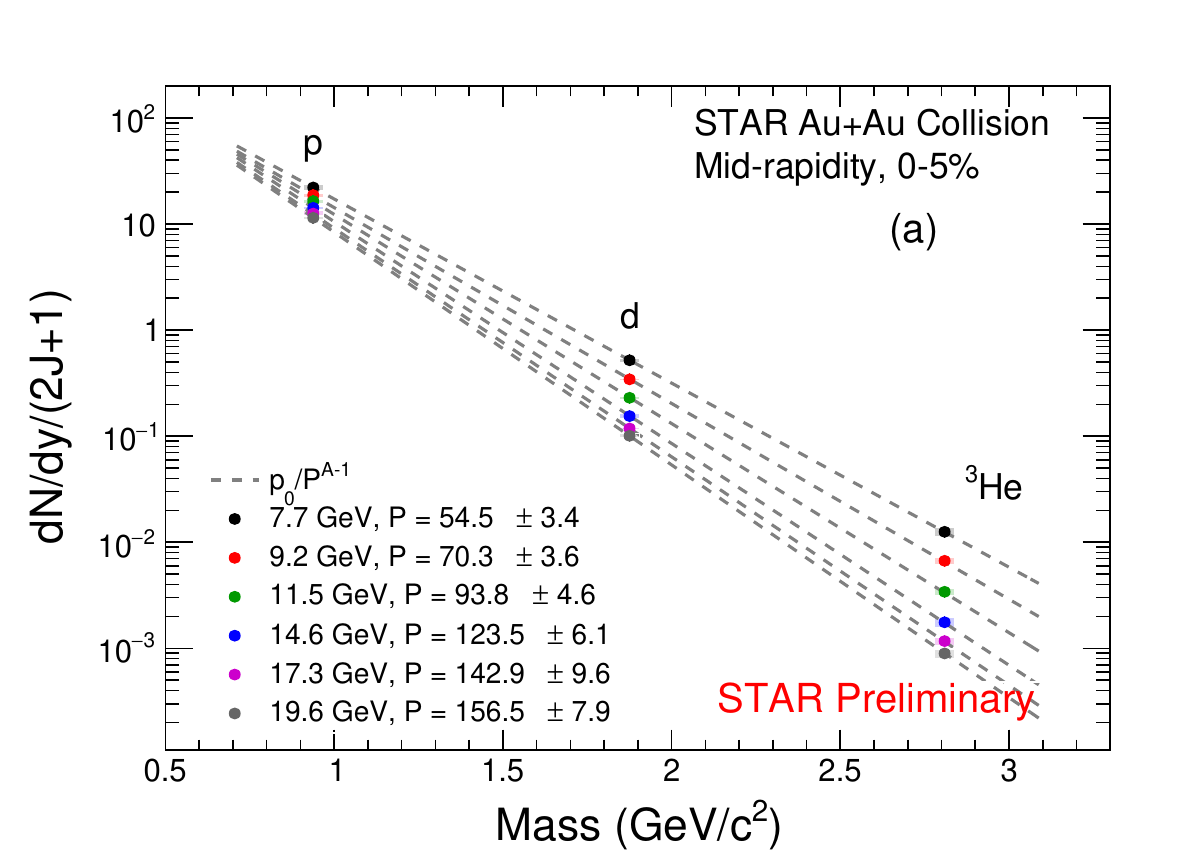}
     \end{subfigure}
     \begin{subfigure}[b]{0.435\textwidth}
          \includegraphics[width=\textwidth]{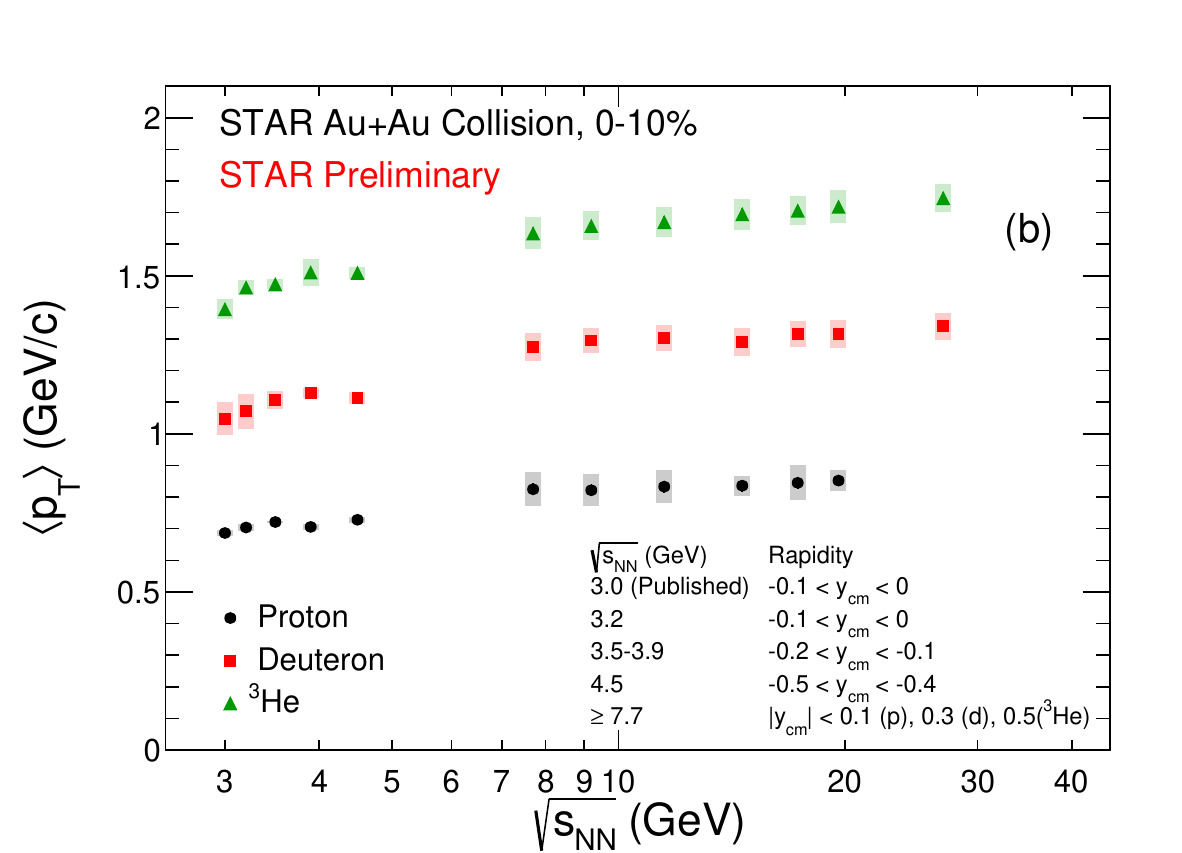}
     \end{subfigure}
     \caption{(a) Mass number ($A$) dependence of $\mathrm{d}N/\mathrm{d}y$, scaled by the spin degeneracy factor $2J + 1$, for primordial protons, deuterons, and $^3$He in Au+Au collisions at BES-II energies of $\sqrt{s_{\rm NN}} = 7.7$--27 GeV. (b) Energy dependence of the average transverse momentum ($\langle p_{\rm T} \rangle$) for protons and light nuclei in 0-10\% central Au+Au collisions at $\sqrt{s_{\rm NN}} = 3$--27 GeV.}
     \label{fig-3}
\end{figure}

The $p_{\rm T}$ spectra are extrapolated to the full range using the Blast-Wave function, the extrapolation uncertainties are estimated by refitting the spectra with a double-exponential function and taking the yield difference from the Blast-Wave fit. $\mathrm{d}N/\mathrm{d}y$ are extracted by integrating over the full $p_{\rm T}$ range. Extended $p_{\rm T}$ coverage reduces the systematic uncertainty in the extrapolated $\mathrm{d}N/\mathrm{d}y$. The yields in Fig.~\ref{fig-3}(a) are scaled by the spin degeneracy factor $2J + 1$. At each energy, particle yields follow $p_0 / P^{A-1}$, where $P = \exp{((m - \mu_B) / T)}$ is the penalty factor, $A$ is the mass number, and $p_0$ is a normalization constant \cite{Zhao:2021dka, 3GeV}. The penalty factor is larger at higher collision energies, reflecting the increased difficulty to form high-mass objects.

Figure \ref{fig-3}(b) presents the average transverse momentum for proton, deuteron, and $\rm ^3He$. For reference, STAR fixed-target measurements in Au+Au collisions at $\sqrt{s_{\mathrm{NN}}} = 3.0$--4.5 GeV \cite{3GeV, Poster} are included. A hint of increase in $\langle p_{\mathrm{T}} \rangle$ is observed. for energies below 4.5 GeV, whereas a flat trend is seen between 7.7 and 27 GeV. The behavior in the intermediate energy region (4.5--7.7 GeV) will be investigated in future studies.

\subsection{Particle Yield Ratios and Coalescence parameters}
\begin{figure}[h]
     \centering
     \begin{subfigure}[b]{0.45\textwidth}
          \includegraphics[width=\textwidth]{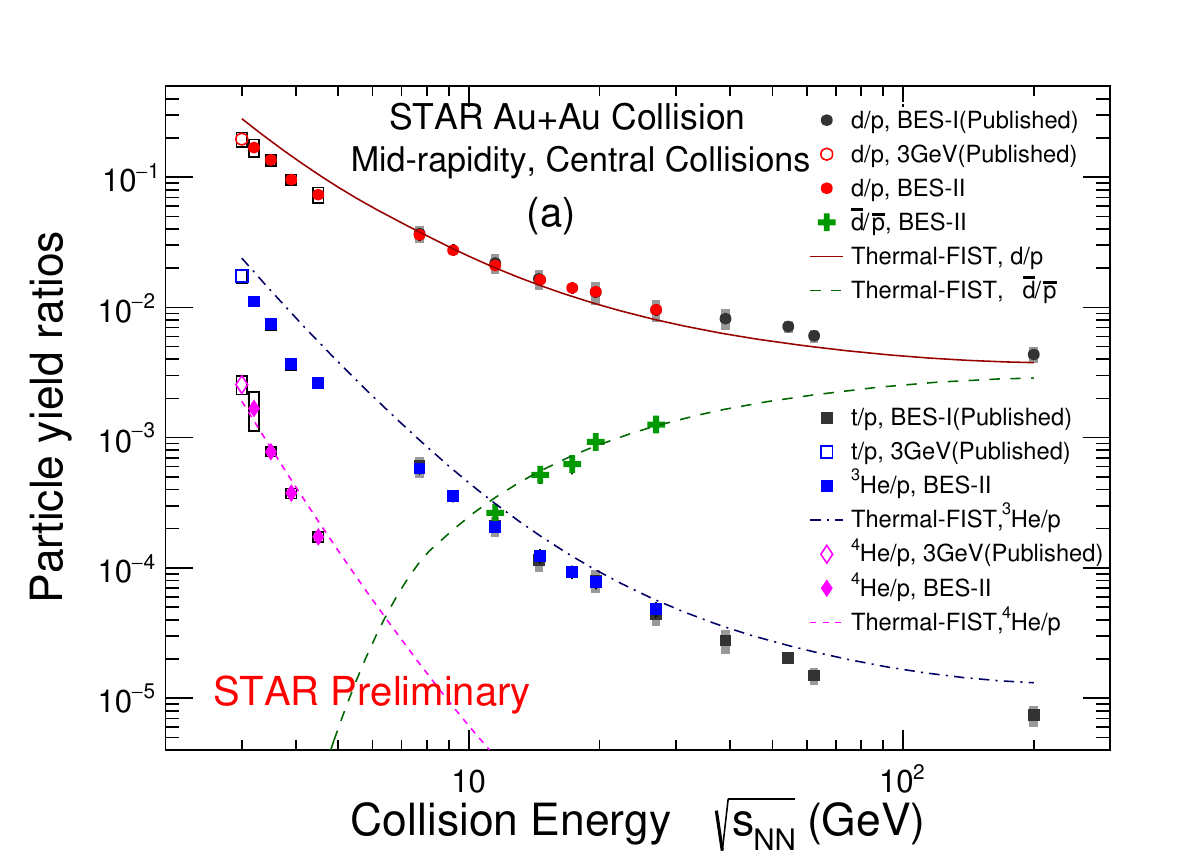}
     \end{subfigure}
     \begin{subfigure}[b]{0.45\textwidth}
          \includegraphics[width=\textwidth]{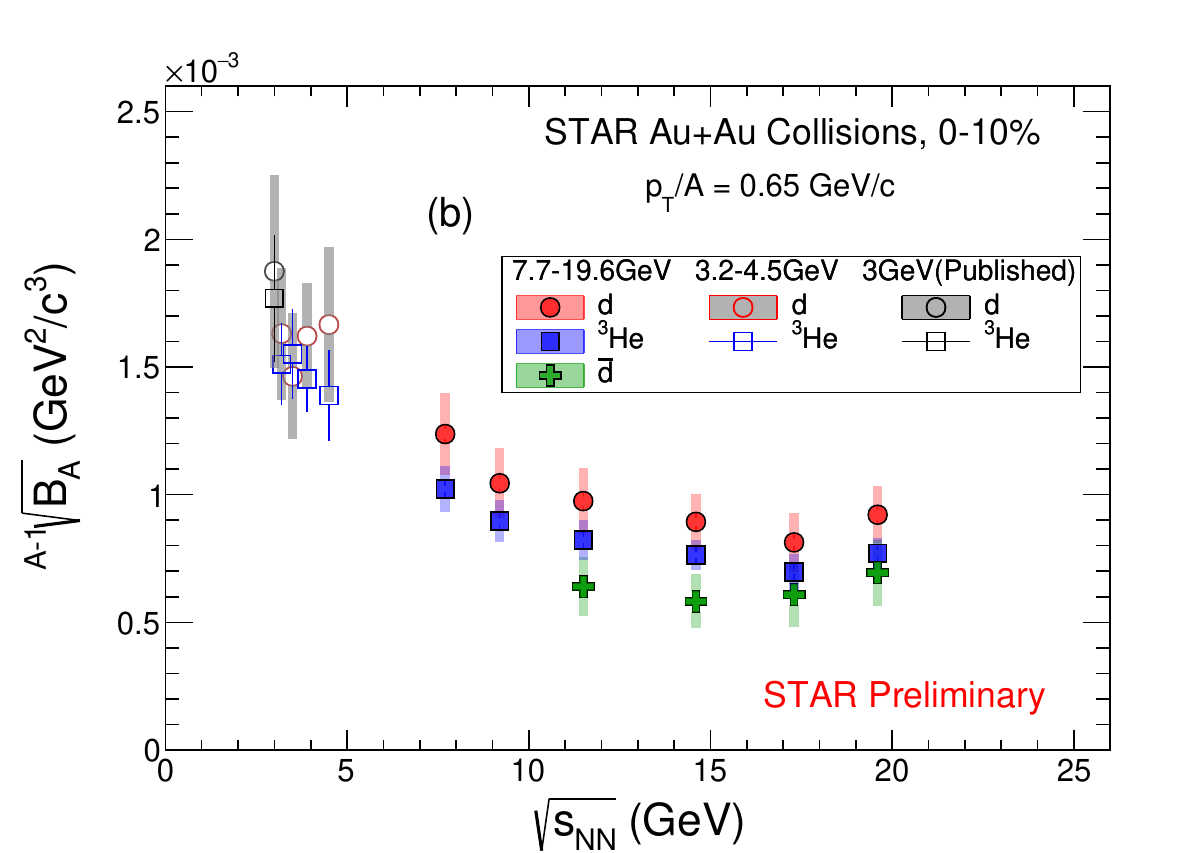}
     \end{subfigure}
     \caption{(a) Light nuclei-to-proton ratios in Au+Au collisions at various energies. Red circles for $\rm d/p$, blue squares for $\rm ^3He/p$, green crosses for $\rm \bar{d}/\bar{p}$, and magenta diamonds for $\rm ^4He/p$. Dashed lines show the thermal model predictions. (b) Energy dependence of $^{A-1}\sqrt{B_A}$ at mid-rapidity in 0--10\% central Au+Au collisions from the STAR BES-II program. The coalescence parameters $B_A$ are evaluated at $p_{\rm T}/A = 0.65$ GeV/$c$. Circles, squares, and crosses represent $B_2({\rm d})$, $B_3({\rm ^3He})$, and $B_2({\rm \bar{d}})$, respectively. 
     Statistical uncertainties are represented by vertical lines, and systematic uncertainties by shaded boxes.}
     \label{fig-4}
\end{figure}

The light nuclei-to-proton ratios, including $\rm d/p$, $\rm ^3He/p$, $\rm ^4He/p$, and $\rm \bar{d}/\bar{p}$, are presented in Fig.~\ref{fig-4}(a). These ratios exhibit a clear energy dependence and qualitatively follow the thermal model. At $\sqrt{s_{\rm NN}} \ge 7.7$ GeV, the $\rm d/p$ and $\rm \bar{d}/\bar{p}$ ratios can be well described by the thermal model including only stable nuclei, while the $\rm ^3He/p$ ratios are overestimated \cite{BESI_t}. This discrepancy may be attributed to hadronic re-scattering effects during the system evolution.

In the coalescence picture,
\begin{equation}
     E_A \frac{\mathrm{d}^3 N_A}{\mathrm{d} p_A^3} = B_A \left( E_p \frac{\mathrm{d}^3 N_p}{\mathrm{d}p_p^3} \right)^Z \left( E_n \frac{\mathrm{d}^3 N_n}{\mathrm{d}p_n^3} \right)^{A-Z} \approx B_A \left( E_p \frac{\mathrm{d}^3 N_p}{\mathrm{d}p_p^3} \right)^A
\end{equation}
 where $A$ is the mass number and $B_A$ is the coalescence parameter. $B_A$ is proportional to $(1 / V_{\rm eff})^{A - 1}$, and quantifies the probability of forming a nucleus from nucleons within the effective emission volume $V_{\rm eff}$ \cite{Coalescence}. In this analysis, neutron yields are estimated using proton yields combined with the $n/p$ ratio from Thermal-FIST. As shown in Fig.~\ref{fig-4}(b), $^{A-1}\sqrt{B_A}$ decreases with collision energy between $\sqrt{s_{\rm NN}} = 3$--11.5 GeV, then remains roughly constant up to 27 GeV, indicating an expansion of the effective volume where nucleons are close enough in phase space to coalesce. The coalescence parameters of d, $\bar{\rm d}$ and $\rm ^3He$ are similar.

\section{Summary}
The $p_{\mathrm{T}}$ spectra and $\mathrm{d}N/\mathrm{d}y$ of light nuclei and their antiparticles at $\sqrt{s_{\mathrm{NN}}} = 7.7$--$27$ GeV from the STAR BES-II program have been presented. The spectra feature reduced systematic uncertainties compared to BES-I results owing to increased statistics and improved detector performance. The ${\rm ^3He} / {\rm p}$ ratios are overestimated by the thermal model. The coalescence parameters $B_A$ decrease with collision energies $\sqrt{s_{\rm NN}} = 3$--11.5 GeV, suggesting an expansion of the effective volume, then roughly stabilize up to 27 GeV. These results provide important insights into the production mechanisms of light nuclei in relativistic heavy-ion collisions.

\section{Acknowledgement}
This work was supported by National Key Research and Development Program of China (No.2022YFA1604900), National Natural Science Foundation of China (No. 12525509 and 12447102).

%
%
%

\end{document}